\pdfoutput=1
\documentclass[epjST]{svjour}
\usepackage[utf8]{inputenc}
\usepackage{amssymb,amsfonts,latexsym,amsmath,times}
\usepackage{bm}
\usepackage{graphicx}
\usepackage{lineno}
\usepackage{geometry}
\usepackage{epstopdf}
\usepackage{subfigure}

\usepackage{multicol}
\usepackage{float}
\usepackage{soul}
\usepackage[titletoc,title]{appendix}
\usepackage[section]{placeins} 
\usepackage[pdftex]{color}

\usepackage{booktabs}
\usepackage{nicefrac}
\usepackage{textcomp}
\usepackage{tabularx}

\begin{document}

\title{Invasion patterns in competitive systems\thanks{Dedicated to Ulrike Feudel on occasion of her 60th birthday}}

\author{Ivo Siekmann\inst{1}\fnmsep\thanks{\email{ivo.siekmann@mathematik.uni-goettingen.de}} \and Michael Bengfort\inst{2} \and Horst Malchow\inst{3}}

\institute{Institute for Mathematical Stochastics, Georg-August-Universit\"at G\"ottingen, Goldschmidtstra\ss{}e 7,\\37077 G\"ottingen, Germany \and Institute of Coastal Research, Helmholtz-Zentrum Geesthacht, Max-Planck-Stra\ss{}e 1,\\21502 Geesthacht, Germany \and Institute of Environmental Systems Research, School of Mathematics\,/\,Computer Science,\\Osnabr\"uck University, Barbarastra\ss{}e 12, 49076 Osnabr\"uck, Germany}

\abstract{
  Stochastic reaction-diffusion equations are a popular modelling
  approach for studying interacting populations in a heterogeneous
  environment under the influence of environmental
  fluctuations. Although the theoretical basis of alternative models
  such as Fokker-Planck diffusion is not less convincing, movement of
  populations is most commonly modelled using the diffusion law due to
  Fick. An interesting feature of Fokker-Planck diffusion is the fact
  that for spatially varying diffusion coefficients the stationary
  solution is not a homogeneous distribution---in contrast to Fick's
  law of diffusion. Instead, concentration accumulates in regions of
  low diffusivity and tends to lower levels for areas of high
  diffusivity. Thus, we may interpret the stationary distribution of
  the Fokker-Planck diffusion as a reflection of different levels of
  habitat quality. Moreover, the most common model for environmental
  fluctuations, linear multiplicative noise, is based on the
  assumption that individuals respond independently to stochastic
  environmental fluctuations. For large population densities the
  assumption of independence is debatable and the model further
  implies that noise intensities can increase to arbitrarily high
  levels. Therefore, instead of the commonly used linear
  multiplicative noise model, we implement environmental variability
  by an alternative nonlinear noise term which never exceeds a certain
  maximum noise intensity. With Fokker-Planck diffusion and the
  nonlinear noise model replacing the classical approaches we
  investigate a simple invasive system based on the Lotka-Volterra
  competition model. We observe that the heterogeneous stationary
  distribution generated by Fokker-Planck diffusion generally
  facilitates the formation of segregated habitats of resident and
  invader. However, this segregation can be broken by nonlinear noise
  leading to coexistence of resident and invader across the whole
  spatial domain.}

\maketitle

\section{Introduction}
\label{sec:intro}

It is needless to give a broad review of the classical publications on
spatial and spatiotemporal pattern formation in non-equilibrium
nonlinear systems. However, on occasion of Ulrike's significant
birthday, one should remember her seminal contributions to the theory
of pattern formation in electrochemical systems from the eighties of
last century \cite{Ebe:83,Feu:84b,Feu:84a}, of course without
forgetting all her impressive later work until today. In that
mentioned early period of her academic career, one of us (H.M.) had
the chance to learn from and to work with Ulrike in Werner Ebeling's
research group at the Sektion Physik of Humboldt-Universit\"at zu
Berlin \cite{Mal:86}. Later on, a theoretical bridging of
electrochemical and ecological diffusive systems was found
\cite{Mal:91,Mal:89b}. Ulrike became interested in ecological and
environmental dynamics as well, and from time to time we manage to
meet and to chat and sometimes even to work, cf. \cite{Ben:14}. { Also
  in the academic career of I.S., Ulrike Feudel has played an
  important role---Ulrike kindly served as an external examiner for
  his PhD defence. He has fond memories of being quizzed on the
  Lotka-Volterra competition model which also plays a central role in
  the study presented here.}

{ Interactions and movements of populations
  $\mathbf{X}(\vec{r},t)=\{X_i(\vec{r},t); ~i=1,2,\ldots,N\}$ in a
  heterogeneous and variable environment are often modelled with
  stochastic reaction-diffusion equations:

  \begin{equation}
    \dfrac{\partial X_i (\vec{r},t)}{\partial t}= \underbrace{f_i(\mathbf{X}(\vec{r}, t))}_{\text{reaction}} + \underbrace{\nabla \cdot \left( - \bm{\mu}(\mathbf{x}, t) X_i(\vec{r}, t) + 
        \nabla [D(\vec{r}, t) X_i(\vec{r}, t)] \right)}_{\text{diffusion}} + \underbrace{g_i(\mathbf{X}(\vec{r}, t))\xi(\vec{r},t)}_{\text{stochastic}} \,, 
    \label{eq:spde}
  \end{equation}

  Here, the reaction terms~$f_i(\mathbf{X}(\vec{r}, t))$ describe the
  interactions between individuals of a population with individuals of
  the same or a different population. This enables us to represent
  processes as diverse as transmission of infectious diseases,
  predator-prey interactions or competition for resources.

  The diffusion term is derived from an underlying stochastic model of
  the movement of individuals. Consider the stochastic differential
  equation for the position~$\mathbf{X}_t \in \mathbb{R}^d$ of a particle that moves
  stochastically in~$d$-dimensional space:
  \begin{equation}
    \label{eq:movesde}
    d \mathbf{X}_t = \bm{\mu}(\mathbf{X}_t, t) dt + \bm{\Sigma}(\mathbf{X}_t, t) d \mathbf{W}_t
  \end{equation}
  The drift coefficients 
  $\bm{\mu}: \mathbb{R}^d \times \mathbb{R}^{+} \to \mathbb{R}^d$ are
  account for deterministic movement of the particle, $d \mathbf{W}_t$
  is a~$d$-dimensional Wiener process and the matrix-valued intensity
  coefficients
  are~$\bm{\Sigma}: \mathbb{R}^d \times \mathbb{R}^{+} \to
  \mathbb{R}^{d\times d}$. An alternative description to the
  SDE~\eqref{eq:movesde} which enables us to calculate the stochastic
  location~$\mathbf{X}_t$ of a particle is the probability
  density~$p(\mathbf{x}, t)$ for finding the particle at
  position~$\mathbf{x}$ at time~$t$. This analogous representation of
  the system is especially suitable for considering a large population
  of particles because in this case, the probability
  density~$p(\mathbf{x}, t)$ can be interpreted as the fraction of
  particles that are expected to be found at a location~$\mathbf{x}$
  at time~$t$. 
  The probability density~$p(\mathbf{x}, t)$ can be shown to satisfy a
  deterministic partial differential equation (PDE), the Kolmogorov
  forward  or Fokker-Planck equation \cite{Kol:31}, \cite[chapter 5]{Oku:01}
  \begin{equation}
    \label{eq:FP}
    \frac{\partial}{\partial t} p(\mathbf{x}, t) = \nabla \cdot \left( - \bm{\mu}(\mathbf{x}, t) p(\mathbf{x}, t) + 
      \nabla [D(\mathbf{x}, t) p(\mathbf{x}, t)] \right)
  \end{equation}
  where~$\nabla$ is the gradient, $\nabla \cdot$ denotes the
  divergence operator and $D(\mathbf{x}, t)$ is related to the
  intensities~$\bm{\Sigma}(\mathbf{x}, t)$ by the standard scalar
  product~$\langle \cdot, \cdot \rangle$:
\begin{equation}
  \label{eq:sigmadiff}
  D_{i,j}(\mathbf{x}, t) = \langle \bm{\Sigma}_{i, \cdot}, \bm{\Sigma}_{j, \cdot} \rangle.
\end{equation}
Now,
choosing~
\begin{equation}
  \label{eq:drift}
  \bm{\mu}(\mathbf{x}, t) := \alpha \nabla D(\mathbf{x}, t), \quad \alpha \in \mathbb{R},  
\end{equation}
we obtain several alternative laws for the dynamics of the probability
distribution~$p(\mathbf{x}, t)$ i.e. the collective movement of the
population. First, it is important to note that in this model
different choices of the parameter~$\alpha$ only have an effect for
spatially varying diffusion coefficients~$D(\mathbf{x},
t)$
. Second, it is easy to see that the stationary solution
of~\eqref{eq:FP} depends on~$\alpha$. For~$\alpha=1
$ an easy
calculation shows that~\eqref{eq:FP} reduces to the well-known
diffusion law due to Fick \cite{Fic:55}
\begin{equation}
  \label{eq:fick}
  \frac{\partial}{\partial t} p(\mathbf{x}, t) = 
  \nabla \cdot \left[  D(\mathbf{x}, t) \nabla p(\mathbf{x}, t) \right]
\end{equation}
which is the diffusion term most commonly used in modelling
applications. From~\eqref{eq:fick} it is clear that the homogeneous
distribution~$p(\mathbf{x}, t) = 1$ is a stationary solution. In
contrast, for~$\alpha=0$ we obtain a law known as Fokker-Planck
diffusion \cite{Fok:14,Pla:17}:
\begin{equation}
  \label{eq:FPdiff}
  \frac{\partial}{\partial t} p(\mathbf{x}, t) = 
  \Delta [D(\mathbf{x}, t) p(\mathbf{x}, t)].
\end{equation}
Here it is obvious that the stationary distribution cannot be
homogeneous for spatially varying diffusion
coefficients~$D(\mathbf{x}, t)$. The reason can be understood by
regarding~\eqref{eq:movesde}. By choosing~$\alpha=0$ the movement of
individual particles is purely stochastic but nevertheless for
spatially inhomogeneous coefficients, the movement is biased towards
directions of larger intensities~$\bm{\Sigma}(\mathbf{x},
t)$. Similarly, we see that in the situation of Fick's law
($\alpha=1$), the drift term $-\nabla D(\mathbf{x}, t)$ opposes the
gradient of~$D(\mathbf{x}, t)$ and in this way seems to exactly
balance the bias of the movement towards larger diffusivities.

For many physicochemical systems, Fick's law is the model of choice
due to the fact that the particles move in a purely passive way which
is consistent with a flux opposed to the concentration gradient. But
for biological populations whose individuals are able to actively
influence their direction of movement, there is no reason to restrict
ourselves to models based on Fick's law. Instead, it seems more
appropriate to start from a general model for stochastic movement such
as~\eqref{eq:FP}. As explained in more detail in \cite{Ben:16b} the
drift term~\eqref{eq:drift} is not only a phenomenological description
but can be interpreted as the ability of an individual to ``sense''
environmental conditions over a distance increasing
with~$0 < \alpha \leq 1$ and ``choose'' its direction of movement
accordingly \cite{Fro:13}. Potapov et al. \cite{Pot:14a}
considered~$\alpha=2$. However, regardless of possible
interpretations of the stochastic movement of individuals underlying a
particular diffusion term, the most important qualitative feature of
the alternative models to Fick's law presented here is the fact that
the stationary distributions of the populations are in most situations
inhomogeneous. Indeed, Bengfort et al. \cite{Ben:16b} showed that
Fokker-Planck diffusion leads to pattern formation for situations
where this would not be expected for Fickian diffusion. 

Whereas Bengfort et al. considered a wide range of deterministic
models, most recently we have also investigated the combined influence
of Fokker-Planck diffusion and stochastic environmental fluctuations
\cite{Ben:16c}. 
As illustrated in \eqref{eq:spde}, by adding a stochastic term,
fluctuations in environmental conditions such as temperature, nutrient
availability etc. over time and/or in space, can be incorporated
without having to represent each source of environmental variability
individually.  Usually, multiplicative Gaussian noise which is
uncorrelated both in space as well as in time is used, i.e. the
standard normally distributed random variable~$\xi_i(\mathbf{x}, t)$
is multiplied by the population density~$X_i$ so that we have
$g_i(\mathbf{X}(\vec{r}, t))=X_i(\vec{r}, t)$.  As explained in more
detail in \cite{Siek:16a}, this particular choice of~$g_i$ can be
related to an individual-based model, the branching process in a
random environment~(BPRE). In a BPRE, the amount of offspring produced
by each individual is modulated by a stochastic process that
represents environmental fluctuations. For large population numbers,
the BPRE can be approximated by a stochastic differential equation
with one term that accounts for demographic stochasticity as well as a
multiplicative noise term~$X_i(\vec{r}, t) \xi_i(\mathbf{x}, t)$
representing the influence of environmental fluctuations. Thus, the
commonly used multiplicative noise model for environmental
stochasticity can be derived from a so-called diffusion approximation
of a BPRE for large populations where demographic stochasticity is
neglected.

That the effect of environmental fluctuations scales with the
population number is essentially due to the fact that environmental
stochasticity is assumed to affect each individual
independently. However, the more the population density increases, the
more likely it seems that individuals are located so close to each
other that instead of responding independently to stochastic
perturbations they are similarly affected. From these considerations it is expected that instead of increasing linearly with the population number, the intensity of the environmental fluctuations saturates, which suggests a model of the following form
\begin{equation}
\label{eq:noise}
g_{i}(\mathbf{X}) = \dfrac{\omega_{i} X_i^m}{\gamma_{i} +
                                                                              \alpha_{i}X_i^n}%
\,.
\end{equation}
which in slightly more general form has recently been proposed by
Siekmann and Malchow \cite{Siek:16a}.  For~$m=n$ the noise
intensity~$g_{ii}$ monotonically tends to a maximal noise
intensity~$\omega_{ii}/\alpha_{ii}$. The half-saturation
constant~$K:=(\gamma_{ii}/\alpha_{ii})^{1/n}$ is the population
density~$X_i$ at which half of the maximal noise intensity is
reached. For~$m<n$ the noise intensity decreases for large population
numbers~$X_i$.  Regardless of the mechanistic interpretation given
above, the most important qualitative difference of~\eqref{eq:noise}
to the multiplicative model is the fact that the noise intensity is
bounded.

The purpose of this study is to examine in more detail the combined
effect of Fokker-Planck diffusion~\eqref{eq:FPdiff} and nonlinear
noise~\eqref{eq:noise}. This is motivated by our most recent work
where we found that by varying the standard approaches for modelling
movement of populations \cite{Ben:16b} or environmental fluctuations
\cite{Siek:16a}, respectively, a wide range of interesting effects
could be observed, even in well-known classical models such as the
Lotka-Volterra competition model. Here, we use the Lotka-Volterra
model for representing an invasive species and investigate which
effect Fokker-Planck diffusion and nonlinear noise have on the success
of the invader. This study continues Bengfort et al. \cite{Ben:16c}
where we combined Fokker-Planck diffusion and the classical
multiplicative environmental noise model with the Lotka-Volterra model
presented here. }

\section{The stochastic competition-diffusion model}

The dynamics of a resident species $X_1$ and an invader $X_2$ is
described by
\begin{align}
\dfrac{\partial X_1}{\partial t}=&(1-X_1)X_1-c_{12}X_1X_2+d_1\nabla^2(X_1D^\ast(x,y))+g_1(X_1)\xi(\vec{r},t)\,,\label{eq:resi}\\
\dfrac{\partial X_2}{\partial t}=&(1-X_2)X_2-c_{21}X_1X_2+d_2\nabla^2 X_2 + g_2(X_2)\xi(\vec{r},t)\,.\label{eq:inva}
\end{align}

\begin{figure}[htbp]
\center
\subfigure[Fokker-Planck]{\includegraphics[width=0.5\textwidth]{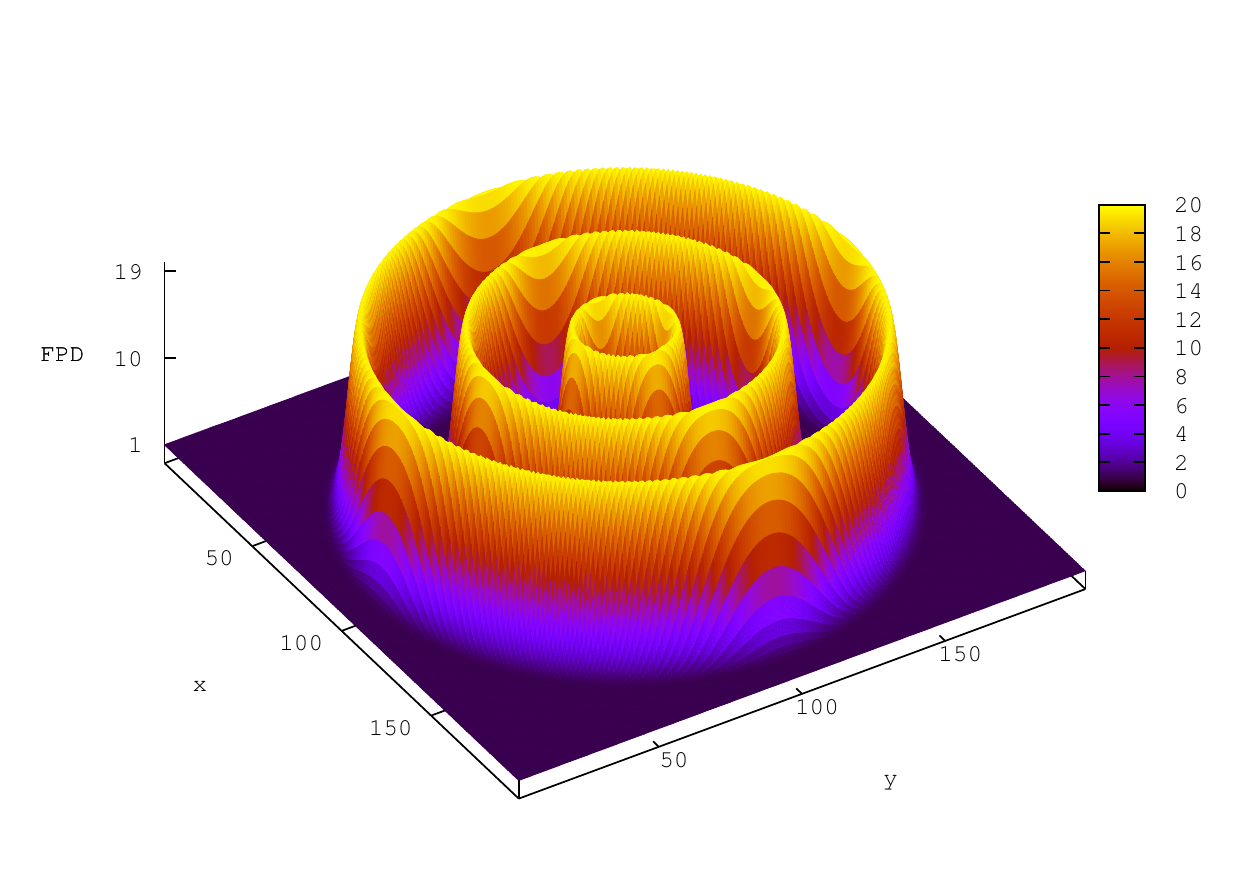}}\hspace*{1cm}
\subfigure[Resident]{\includegraphics[width=0.5\textwidth]{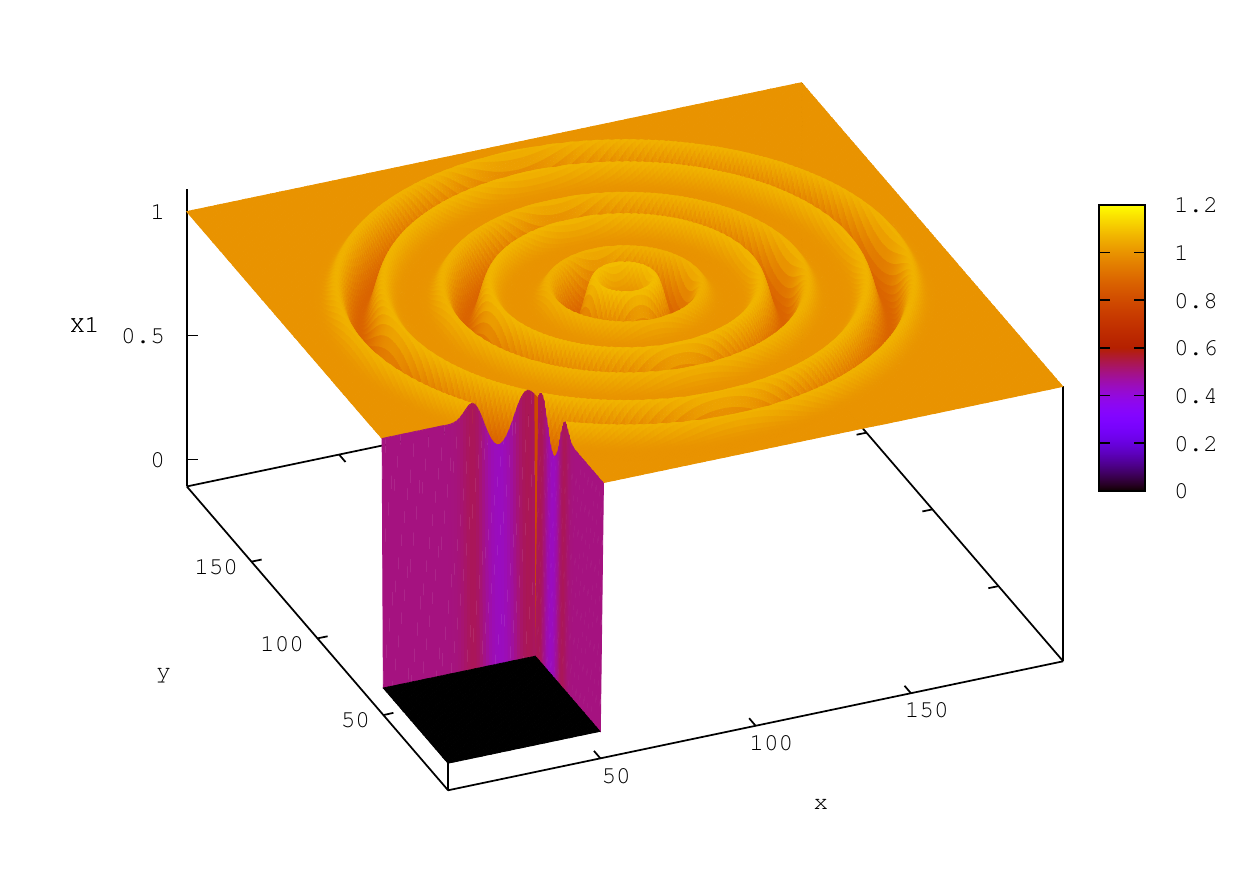}}
\caption{Profile of Fokker-Planck diffusion and initial setting for residents}
\label{fig:0000}
\end{figure}

The spatial dependency of the resident's diffusivity is chosen as
\begin{equation}
\label{eq:diffu}
D^\ast(x,y)=D_0+\left\{\begin{array}{lc} a\left(\sin(\sqrt{x^2+y^2})\right)^k & \text{if }\sqrt{x^2+y^2}<3\pi\,,\\
			a\left(\sin(3\pi)\right)^k & \text{else\,.}\end{array}\right.
\end{equation}
Here, the parameter $k$ is an even number witch controls the steepness
of $D^\ast$. Throughout this paper we will use the parameters 
\[
  D_0=1\,, ~a=19\,, ~k=8
\]
This functional form of the diffusivity~$D^\ast(x,y)$,
see Figure~\ref{fig:0000} for a plot, is meant to represent the
resident's varying levels of preference for different areas of the
spatial domain. The coefficient~$D^\ast(x,y)$ can be regarded as being
inversely proportional to the resident's preference for a particular
location~$(x,y)$. Namely, the lower~$D^\ast(x,y)$, the lower the
tendency to leave~$(x,y)$ which can be interpreted as a high level of
preference. How these preferences for different parts of the habitat
affect the stationary distribution of the resident is fundamentally
different for Fickian diffusion and Fokker-Planck diffusion. For
Fickian diffusion the spatially heterogeneous diffusion
coefficient~$D^\ast(x,y)$ only affects the transient dynamics of the
resident because the stationary distribution is always homogeneous,
regardless of the particular functional form of~$D^\ast(x,y)$. In
contrast, for Fokker-Planck diffusion, the stationary solution is
approximately inversely proportional to~$D^\ast(x,y)$ which provides
us with a simple model for a fragmented habitat that mimics the
resident's levels of preference. In the absence of the invader~$X_2$,
the resident~$X_1$ tends to the distribution shown in the right panel
of Figure~\ref{fig:0000}---note that for the initial condition we have
always set the resident's population to zero within a square with a
side length of 50 length units in order to mimic the onset of a
biological invasion.

We ensure that resident and invader do not differ in competitive
strength by letting the competition parameters coincide
\[
c_{12}=c_{21}=1.2
\]
Because both~$c_{12}$ and~$c_{21}$ exceed unity, the system is in the
bistable parameter range i.e. in the absence of diffusion or noise the
competitor with the larger initial density will drive its opponent to
extinction. For spatially extended systems, Malchow et al. showed that
survival depends on the ratio of the diffusion coefficients of invader
and resident---in general, the competitor with the larger diffusion
coefficient prevails \cite{Mal:11}. Because the spatially varying
diffusion coefficient~$d_1\cdot D^\ast(x,y)$ of the resident is larger
than the constant diffusion coefficient~$d_2$ of the invader in some
areas and smaller in others, it is expected from Malchow et
al. \cite{Mal:11} that the spatial domain becomes segregated. Whereas
in some regions invasion is successful due to relatively low
diffusivity of the resident, other areas act as barriers for invasion
where the resident's diffusivity is relatively high. We will
vary~$d_1$ and~$d_2$ in order to explore the effect of different ratios
of the local diffusion coefficients.

We will investigate this situation under the influence of nonlinear
noise~\eqref{eq:noise}, thereby extending our previous study where we
applied the standard model of multiplicative noise \cite{Ben:16c}. 
Thus, throughout the manuscript the noise intensities~$g_1$ and~$g_2$
are of the form~\eqref{eq:noise} for which we always choose
\[
\alpha_1=\alpha_2=0.1.
\]
The parameters~$\omega_i$ and~$\gamma_i$ will be varied in order to
study different dependencies of the noise intensity on the population
densities~$X_i$.

\begin{figure}[htbp]
\center
\subfigure[Fick]{\includegraphics[width=0.175\textwidth]{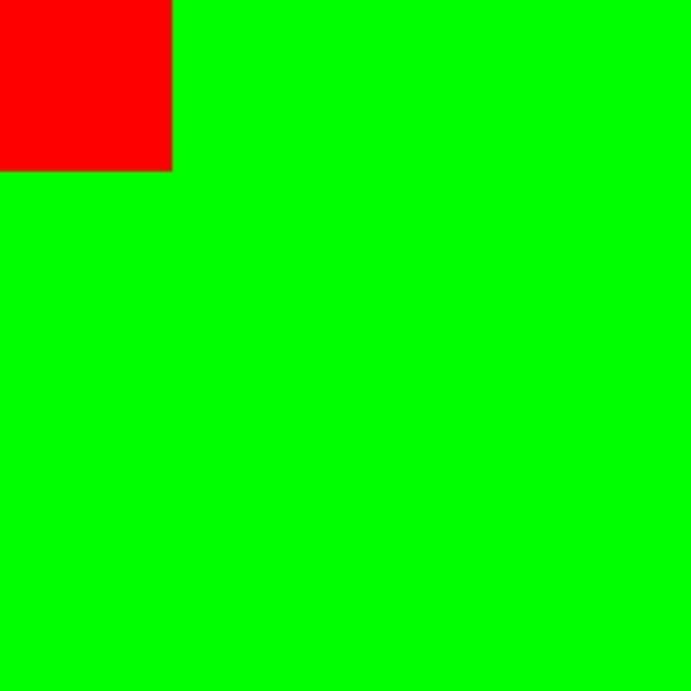}}\hspace*{1cm}
\subfigure[Fokker-Planck]{\includegraphics[width=0.175\textwidth]{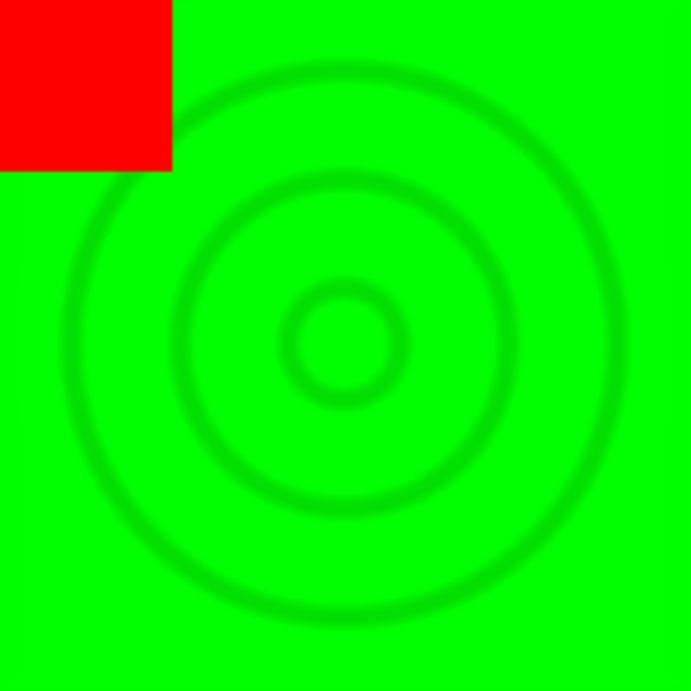}}
\caption{2D projection of initial settings for population densities
  (green = resident, red = invader), cf. Figure~\ref{fig:0000}.}
\label{fig:000}
\end{figure}

The initial condition for a spatial grid of $200 \times 200$ grid
cells is indicated in Figures~\ref{fig:0000} and~\ref{fig:000}, the
invader is set to zero and the resident is initialised with the
spatially heterogeneous stationary distribution in the whole spatial
domain except for a patch of $50 \times 50$ grid cells in the upper
left. Here, the resident is set to zero whereas the invader is set to
its carrying capacity~1. An exception is one simulation (see
Figures~\ref{fig:009} and~\ref{fig:010}) where we compare Fick's law
with Fokker-Planck diffusion where the resident is initialised with a
spatially homogeneous distribution.

We numerically solve~\eqref{eq:resi}, \eqref{eq:inva} using an
alternating direction implicit (ADI) scheme for efficiently
implementing the Crank-Nicholson method \cite{Cra:47} as described
previously \cite{Ben:16c}. The stochastic terms for which we use the
Stratonovich interpretations are numerically integrated with the
derivative-free Milstein method \cite{Klo:99,Mil:95} as explained in
\cite{Siek:16a}. The temporal and spatial step sizes (in
non-dimensional units) are usually
\[
  h_t=0.02, h_x=h_y=15
\]
except for Figures~\ref{fig:009} and~\ref{fig:010} where a smaller
temporal step width of~$h_t=0.002$ was required. 

\section{Numerical simulations and results}

{ With the parameters described in the previous section we
  now investigate the success of invasion depending on various choices
  of the scaling parameters~$d_1$ and~$d_2$ of the resident's and
  invader's diffusivities as well as different parameters~$\omega_i$
  and $\gamma_i$ for the noise. As mentioned above, an initial patch
  of invaders of $50 \times50$ grid cells in size, i.e., 6.25\% of the
  total model area, is placed in a corner of the spatial domain,
  cf. Figures~\ref{fig:0000} and~\ref{fig:000}. Parameters for the
  simulations presented here are summarised in
  Table~\ref{tab:parameters}.

  \begin{table}[htbp]
    \centering
\begin{tabularx}{\textwidth}{|X|X||X|X|X|X|X|X|X|X|}
\hline
$m$	&$n$	&$d_1$	&$d_2$	&a   &$\omega_1$&$\omega_2$&$\gamma_1$&$\gamma_2$& Fig.\\
\hline\hline
1	&1	&5	&25	&19	&0.1	&0.2		&1.0	&1.0	& \ref{fig:001}\\
1	&1	&25	&5	&19	&0.1	&0.2		&1.0	&1.0	& \ref{fig:002}\\
1	&1	&25	&12.5	&19	&0.1	&0.2		&1.0	&1.0& \ref{fig:003}\\
1	&1	&25	&12.5	&19	&0.1	&0.4		&1.0	&1.0& \ref{fig:005}\\
\hline\hline
1	&2	&25	&12.5	&19	&0.1	&0.4		&1.0	&1.0& \ref{fig:008}\\
\hline\hline
1	&2	&25	&12.5	&0	&0.1	&0.4		&1.0	&0.1& \ref{fig:009}\\
\hline
1	&2	&25	&12.5	&19	&0.1	&0.4		&1.0	&0.1& \ref{fig:010}\\
\hline
\end{tabularx} \caption{Parameter values for the simulations presented
  in Figures~\ref{fig:001}-\ref{fig:010}.}
    \label{tab:parameters}
  \end{table}

  The influence of the ratio of
  ~$d_1$ and~$d_2$ is demonstrated in
  Figures~\ref{fig:001}-\ref{fig:003}. For Figures~\ref{fig:001}
  and~\ref{fig:002}, the only difference between both parameter sets
  is that~$d_1$ and~$d_2$ are swapped.  Consistent with Malchow et
  al. \cite{Mal:11} mentioned in the previous section if~$d_1$ is low
  compared to~$d_2$, the invasion is successful (Figure~\ref{fig:001})
  whereas it fails for~$d_1$ much larger than~$d_2$
  (Figure~\ref{fig:002}), although the invader initially seems to be
  able to enter the realm of the resident ($t=100$). Interestingly,
  Figure~\ref{fig:001} shows that even in the case of successful
  invasion, the invader is not able to overcome all barriers created
  by large diffusivities of the resident so that the resident prevails
  in the centre of the spatial domain. The result is a segregation of
  the spatial domain in one habitat dominated by the invader and one
  habitat dominated by the resident.

\begin{figure}[htbp]
\center
\subfigure[$t=500$]{\includegraphics[width=0.175\textwidth]{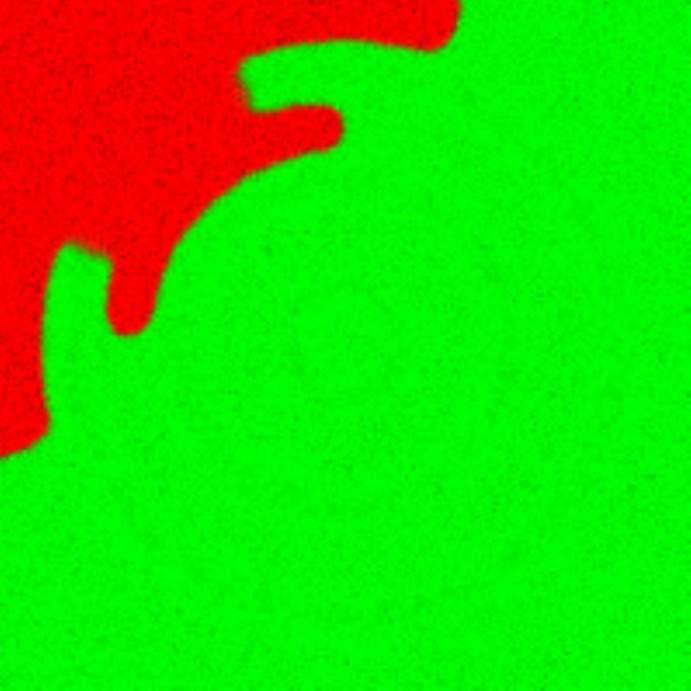}}\hfill
\subfigure[$11000$]{\includegraphics[width=0.175\textwidth]{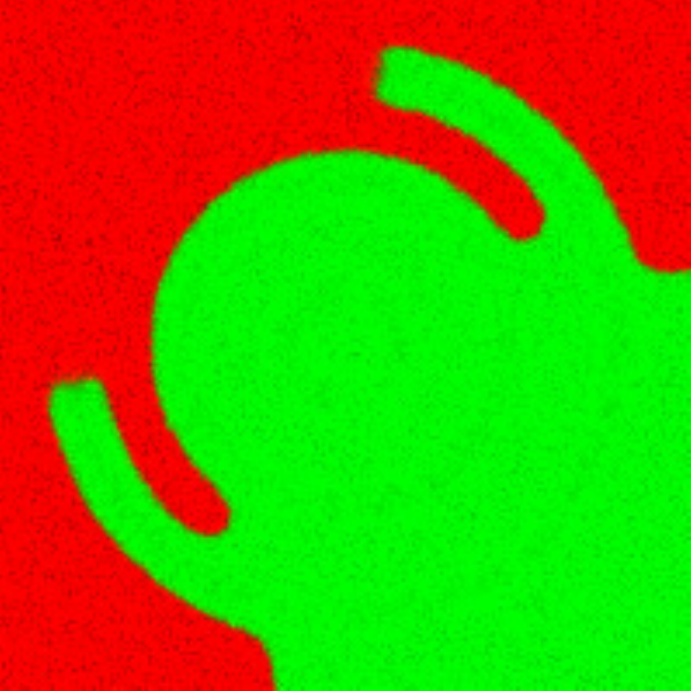}}\hfill
\subfigure[$17000$]{\includegraphics[width=0.175\textwidth]{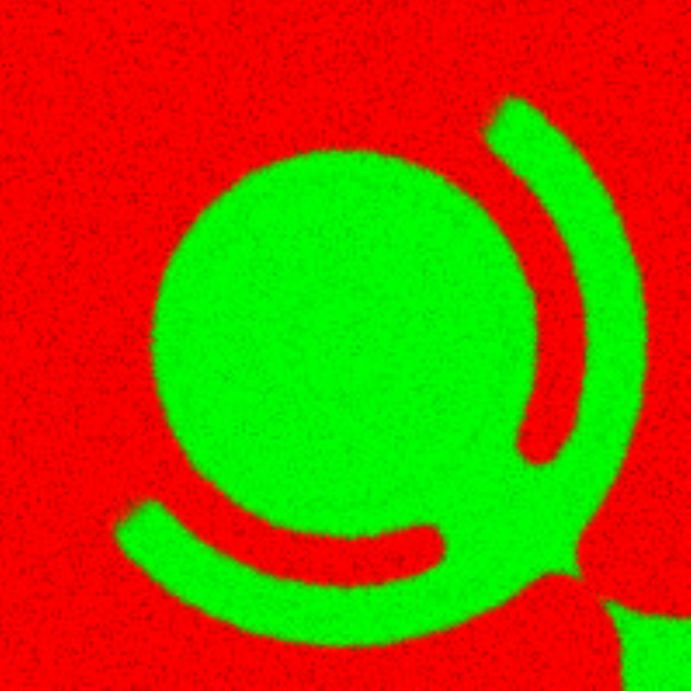}}\hfill
\subfigure[$23000$]{\includegraphics[width=0.175\textwidth]{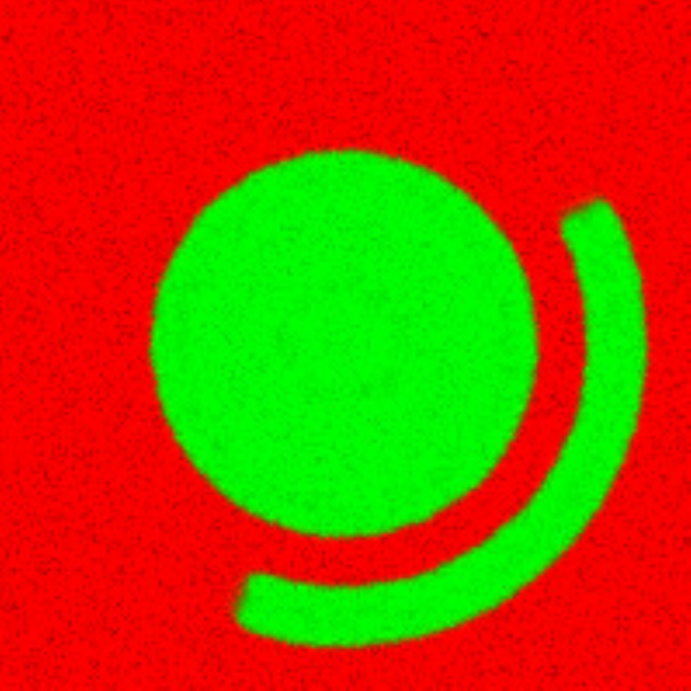}}\hfill
\subfigure[$29000$]{\includegraphics[width=0.175\textwidth]{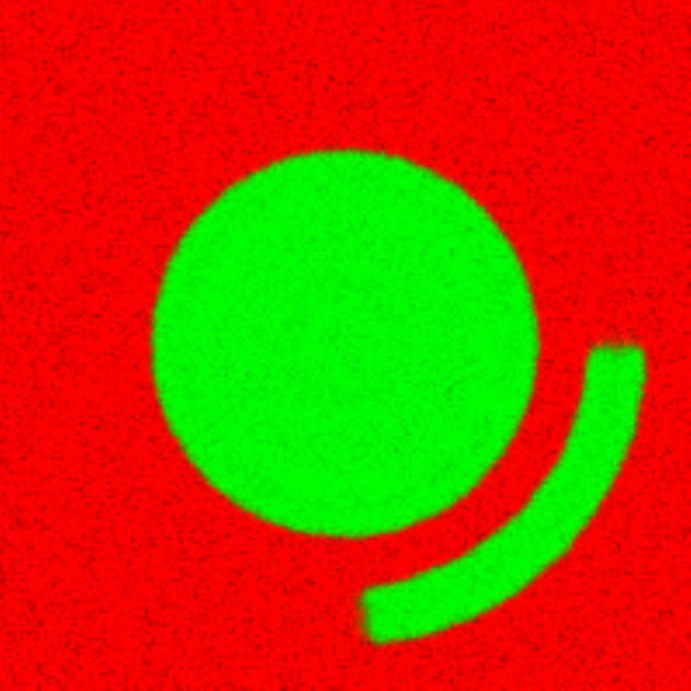}}\hfill
\caption{Successful invasion for a low diffusivity of the resident
  compared with the invader.}
\label{fig:001}
\end{figure}

\begin{figure}[htbp]
\center
\subfigure[$t=100$]{\includegraphics[width=0.175\textwidth]{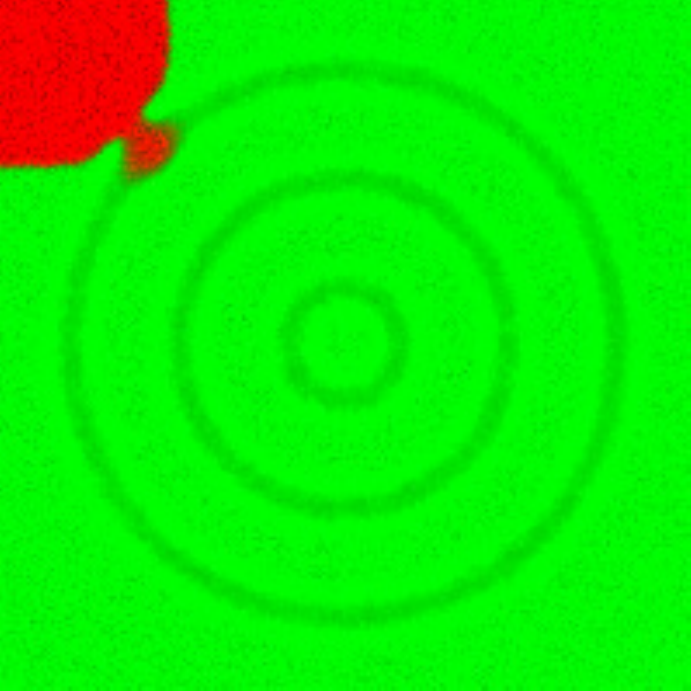}}\hfill
\subfigure[$500$]{\includegraphics[width=0.175\textwidth]{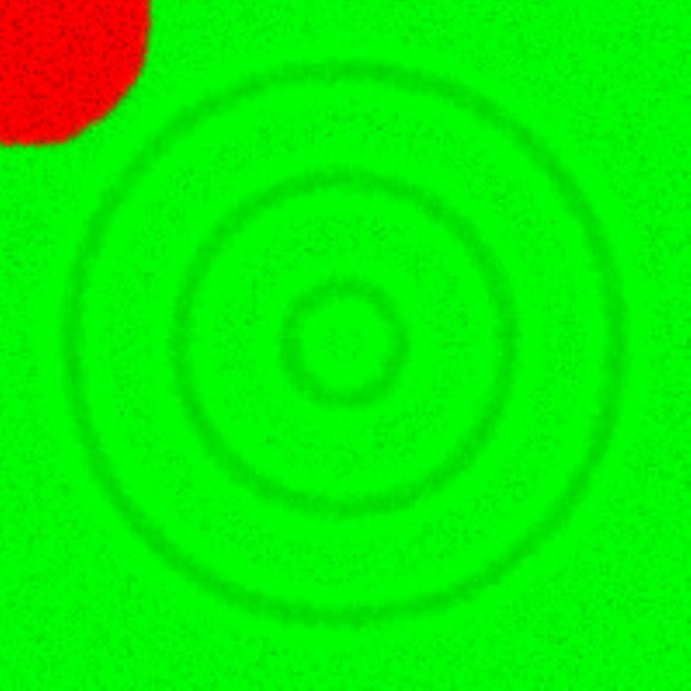}}\hfill
\subfigure[$1000$]{\includegraphics[width=0.175\textwidth]{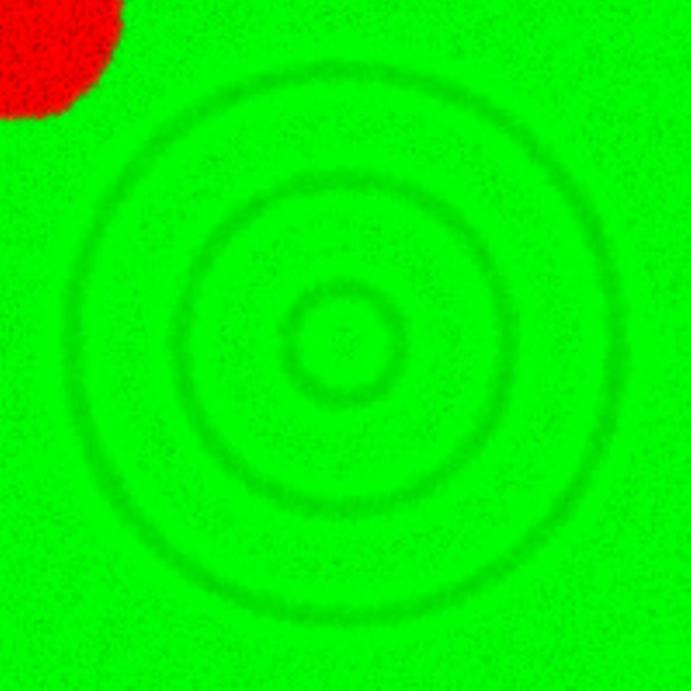}}\hfill
\subfigure[$2000$]{\includegraphics[width=0.175\textwidth]{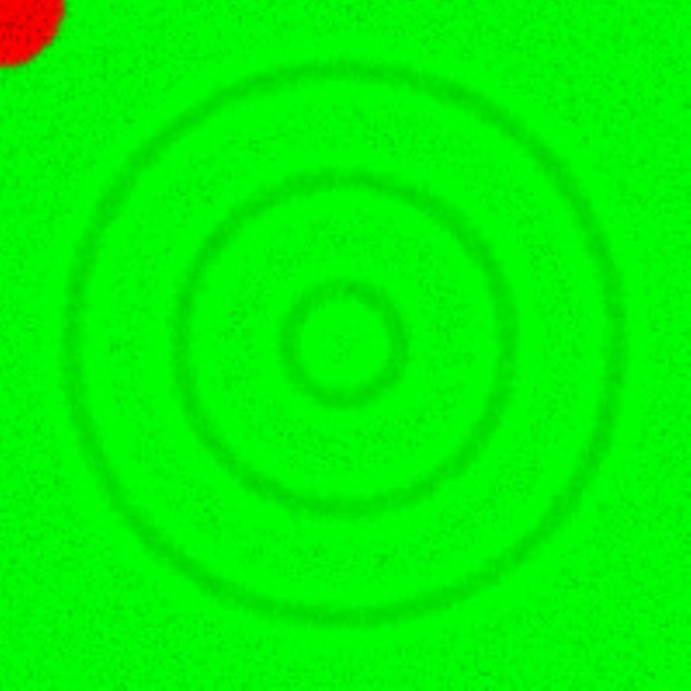}}\hfill
\subfigure[$3000$]{\includegraphics[width=0.175\textwidth]{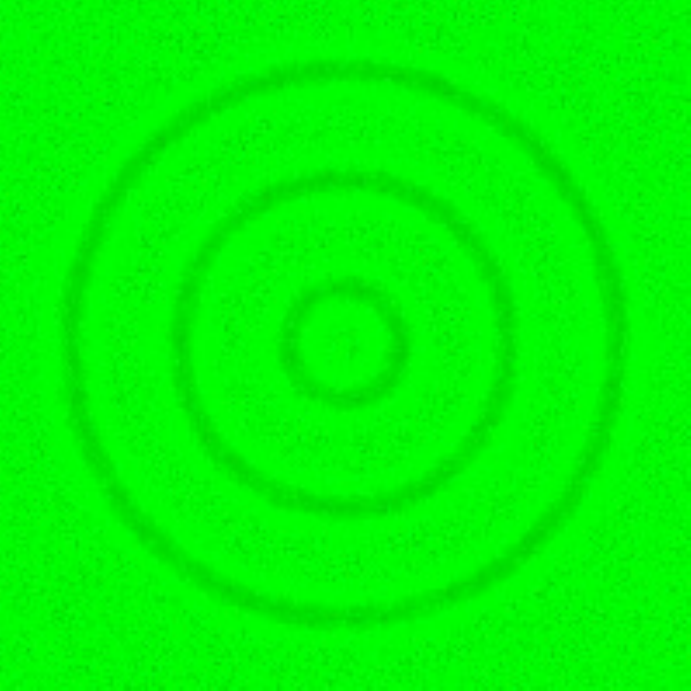}}\hfill
\caption{Invasion fails if the invader's diffusivity is low compared with the resident.}
\label{fig:002}
\end{figure}

  For Figure~\ref{fig:003}, the diffusivity~$d_2$ has been
  increased. In contrast to Figure~\ref{fig:002} the invader not only
  manages to proceed into the territory of the resident but succeeds
  in establishing itself along a ring where the stationary population
  of the resident is at a low level. Very slowly, the invader manages
  to displace the resident from this ring without being able to occupy
  any other region in the spatial domain. Thus, like in
  Figure~\ref{fig:001} we again end up with spatially segregated
  habitats but here the invader is confined between two regions
  occupied by the resident.

\begin{figure}[htbp]
\center
\subfigure[$t=5000$]{\includegraphics[width=0.175\textwidth]{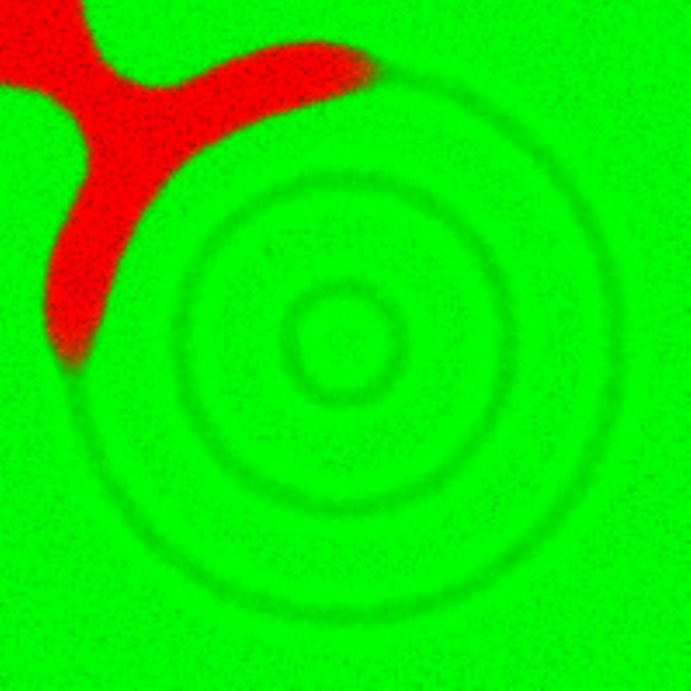}}\hfill
\subfigure[$7000$]{\includegraphics[width=0.175\textwidth]{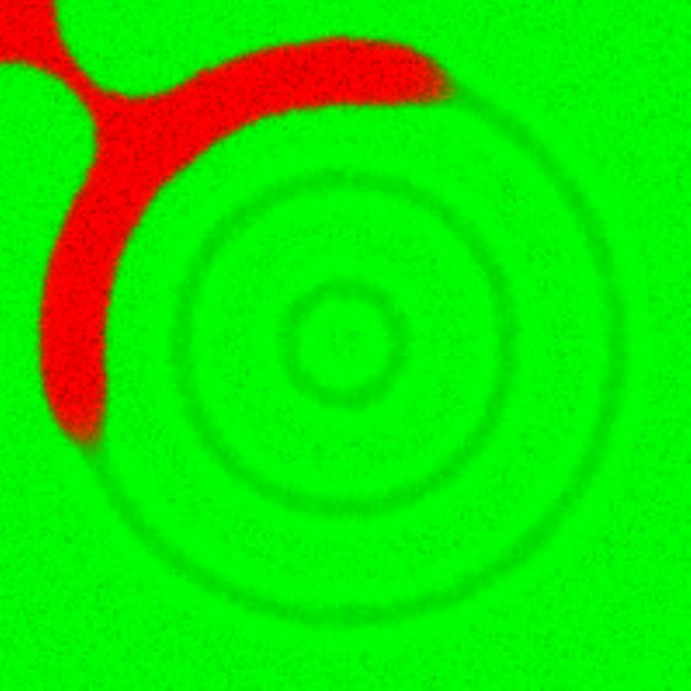}}\hfill
\subfigure[$12000$]{\includegraphics[width=0.175\textwidth]{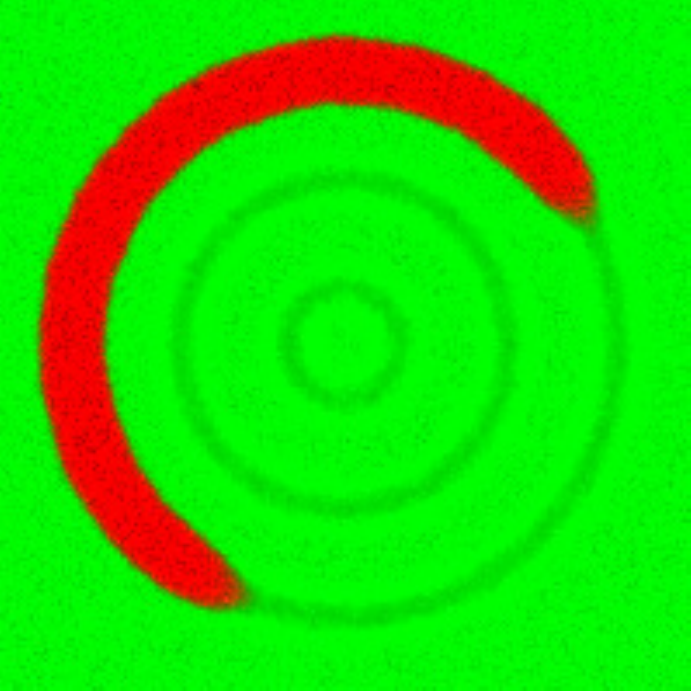}}\hfill
\subfigure[$17000$]{\includegraphics[width=0.175\textwidth]{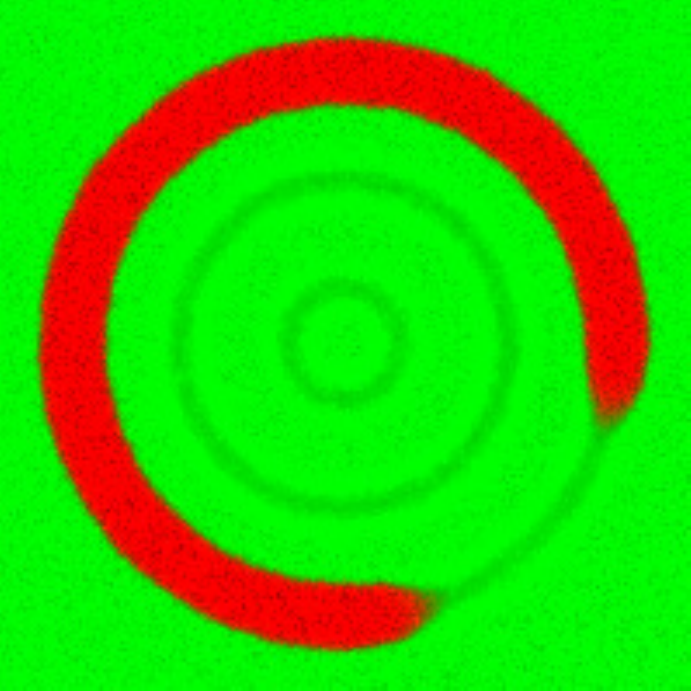}}\hfill
\subfigure[$22000$]{\includegraphics[width=0.175\textwidth]{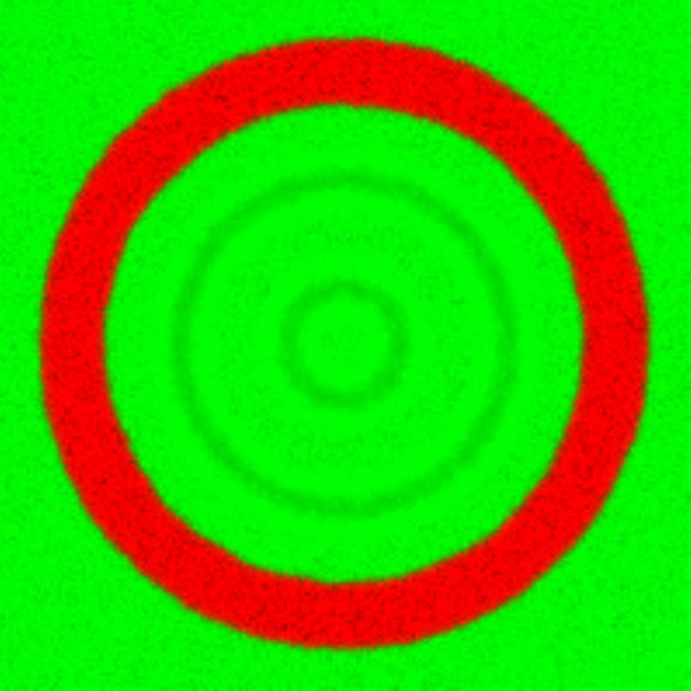}}\hfill
\caption{For an intermediate level of invader's diffusivity the
  invader manages to occupy a less preferred area of the resident's
  original habitat.}
\label{fig:003}
\end{figure}

  Increasing the noise intensity of the invader increases its ability
  to cross invasion barriers caused by large values of the diffusion
  coefficient~$D^{\ast}(x,y)$ of the resident. Consistently, in
  Figure~\ref{fig:005} we observe a similar situation as in our
  initial simulation (Figure~\ref{fig:001}) with low diffusivity of
  the resident in comparison with the invader. Again, the invader is
  able to conquer a habitat that extends to the boundary of the
  spatial domain instead of remaining constrained to a ring. 

\begin{figure}[htbp]
\center
\subfigure[$t=1000$]{\includegraphics[width=0.175\textwidth]{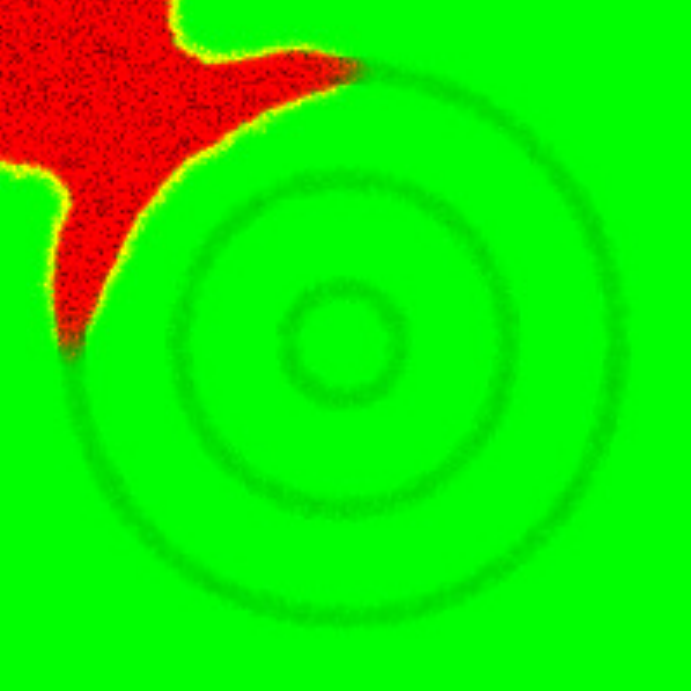}}\hfill
\subfigure[$5000$]{\includegraphics[width=0.175\textwidth]{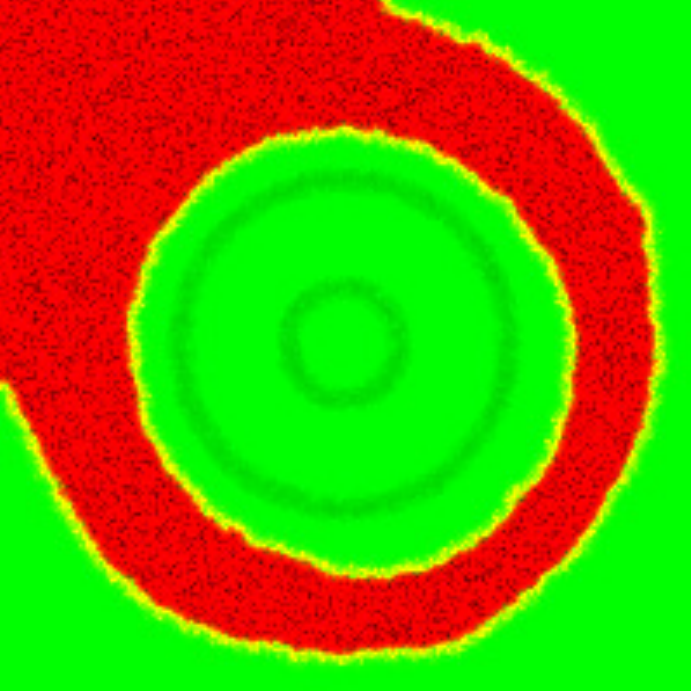}}\hfill
\subfigure[$10000$]{\includegraphics[width=0.175\textwidth]{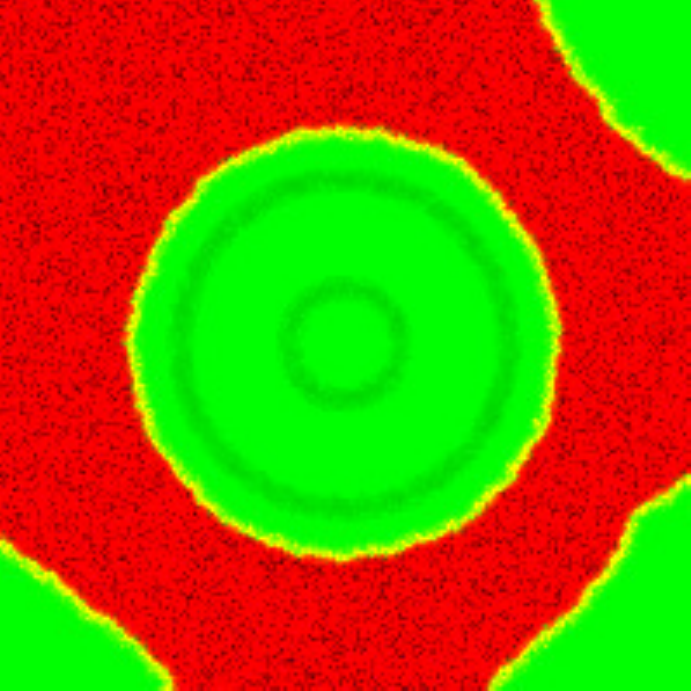}}\hfill
\subfigure[$15000$]{\includegraphics[width=0.175\textwidth]{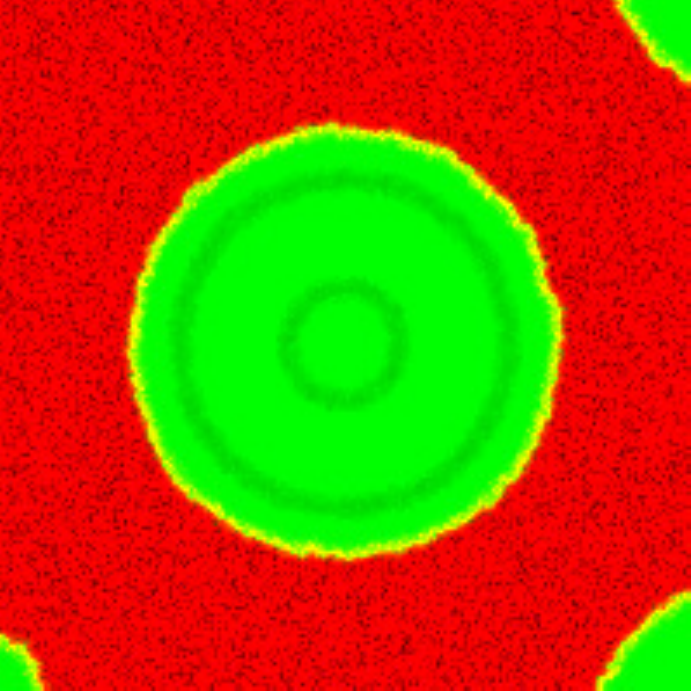}}\hfill
\subfigure[$20000$]{\includegraphics[width=0.175\textwidth]{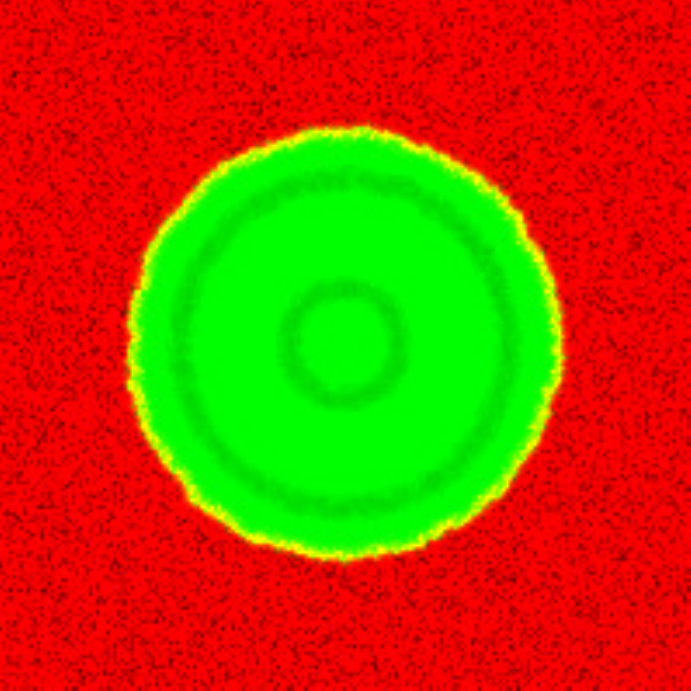}}\hfill
\caption{Increasing the invader's noise intensity~$\omega_2$ has a
  similar effect as increasing diffusivity, in comparison with
  Figure~\ref{fig:003} the invader establishes itself in a larger
  spatial domain.}
\label{fig:005}
\end{figure}

  In Figure~\ref{fig:008} we see that this result is hardly unchanged
  if we replace the monotonously increasing noise by terms that
  initially increase and then decrease for large densities. This is
  achieved by choosing a larger exponent in the denominator
  of~\eqref{eq:noise} i.e. $m<n$. The resulting pattern strongly
  resembles Figure~\ref{fig:005}, also note that the dynamics even
  develops on a similar time scale.

\begin{figure}[htbp]
\center
\subfigure[$t=3000$]{\includegraphics[width=0.175\textwidth]{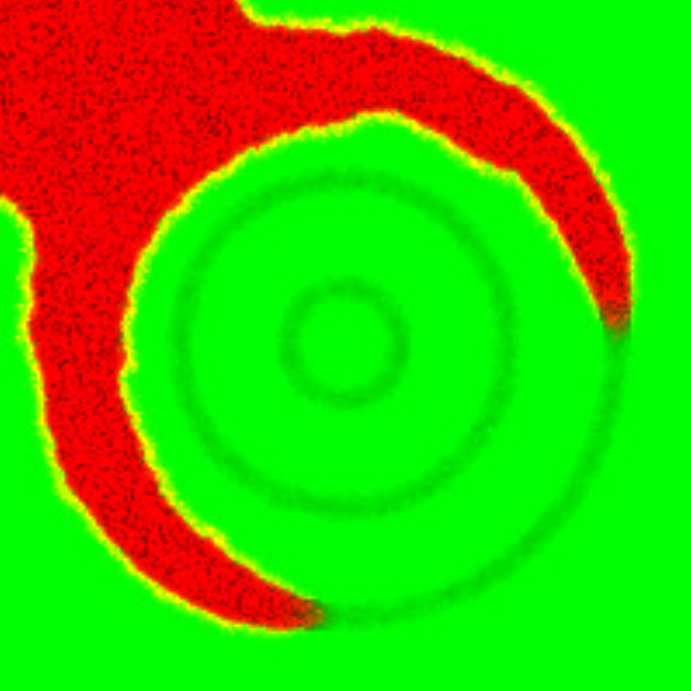}}\hfill
\subfigure[$6000$]{\includegraphics[width=0.175\textwidth]{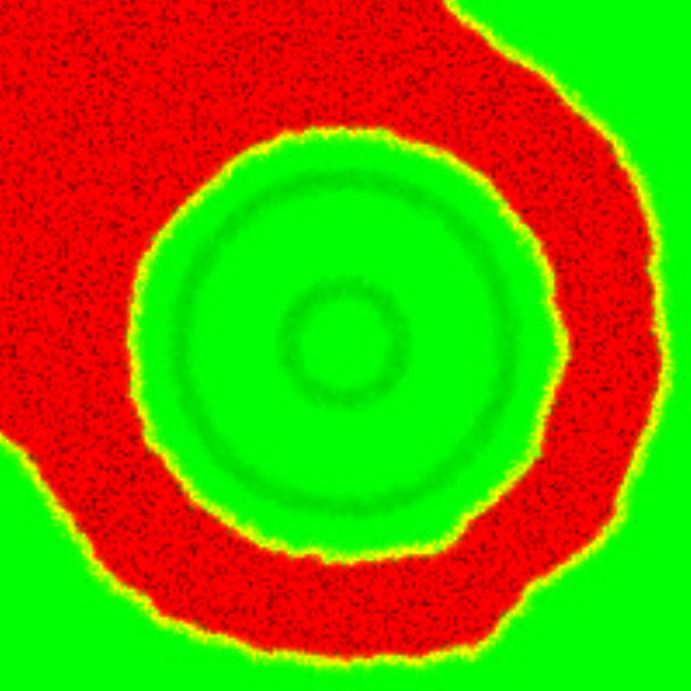}}\hfill
\subfigure[$9000$]{\includegraphics[width=0.175\textwidth]{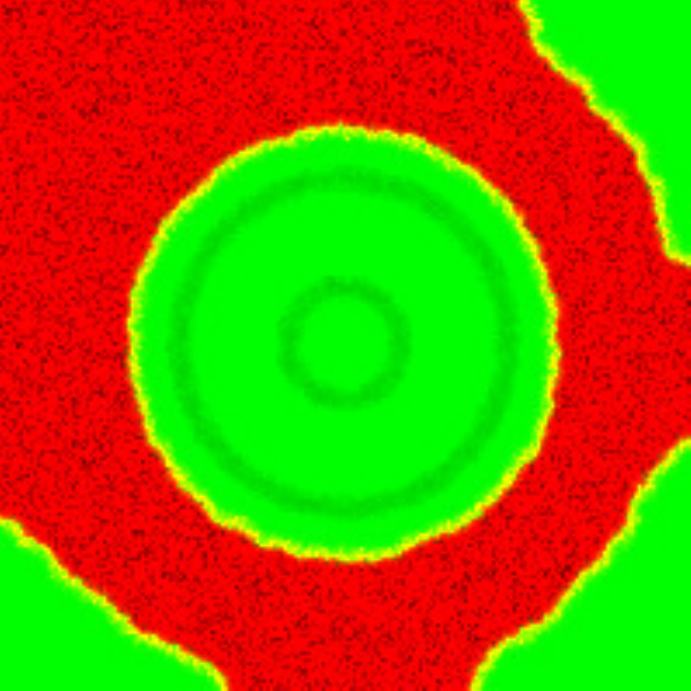}}\hfill
\subfigure[$12000$]{\includegraphics[width=0.175\textwidth]{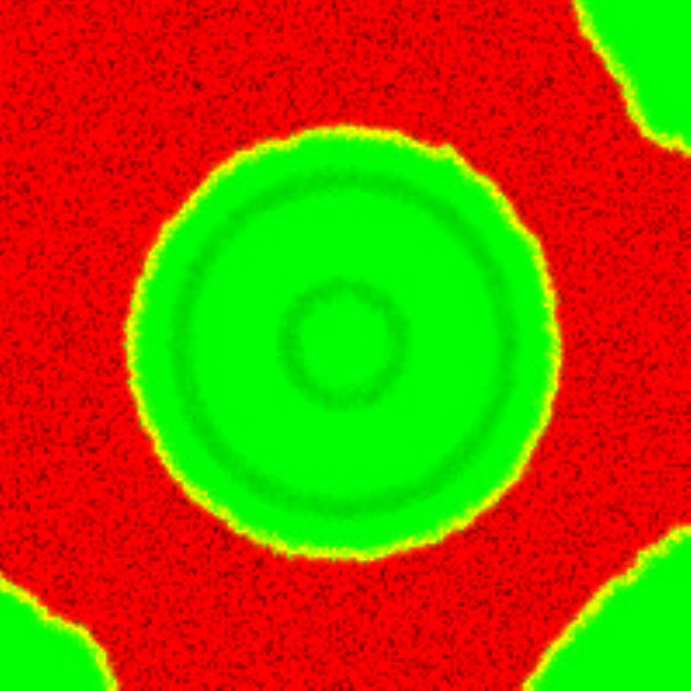}}\hfill
\subfigure[$20000$]{\includegraphics[width=0.175\textwidth]{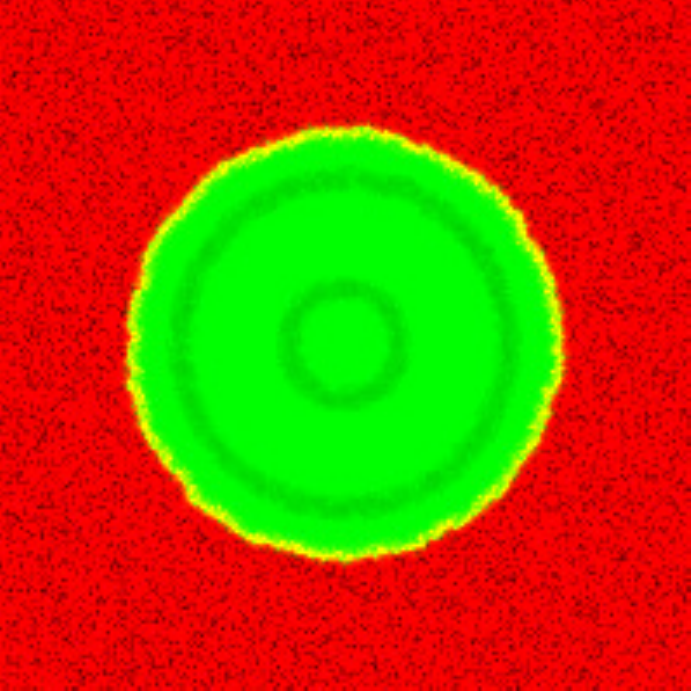}}\hfill
\caption{By choosing~$n=2$ the noise intensity decreases for large
  population densities rather than saturating at a maximal noise
  intensity. However, in the absence of further changes, the
  differences to Figure~\ref{fig:005} for $m=n=1$ are minor.}
\label{fig:008}
\end{figure}

  For this form of the noise term even coexistence of invader and
  resident within the whole spatial domain rather than the emergence
  of separate habitats is possible. The fact that the coexistence
  pattern can be observed both for Fickian as well as Fokker-Planck
  diffusion shows that this is an effect mostly mediated by the noise
  term. In comparison with the previous simulation
  (Figure~\ref{fig:008}) we shift the half-saturation constant of the
  invader to lower population densities by decreasing~$\gamma_2$. This
  has the effect of a steeper increase of the noise
  term~\eqref{eq:noise}. In Figure~\ref{fig:009} we show that for
  Fickian diffusion both populations quickly mix. The initial invader
  patch develops into a mixed front consisting of both resident and
  invader that replaces the area formerly occupied only by the
  resident. 

\begin{figure}[htbp]
\center
\subfigure[$t=10$]{\includegraphics[width=0.175\textwidth]{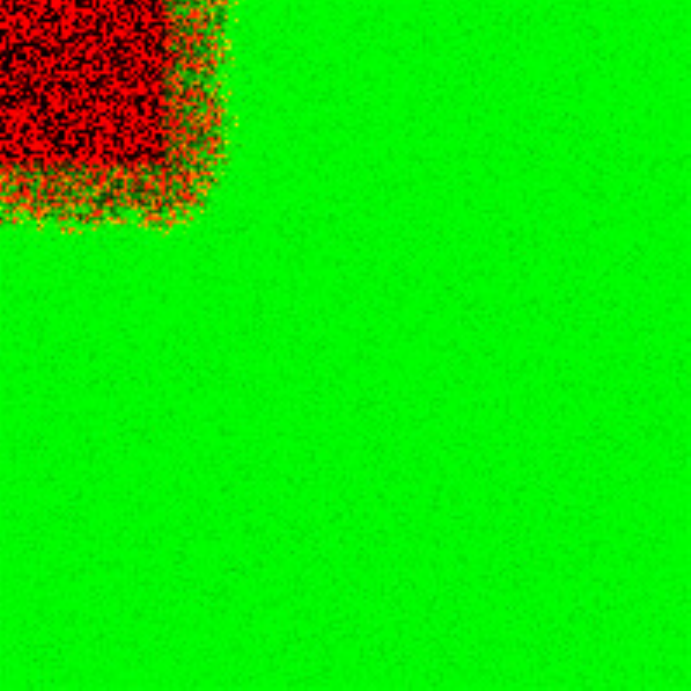}}\hfill
\subfigure[$50$]{\includegraphics[width=0.175\textwidth]{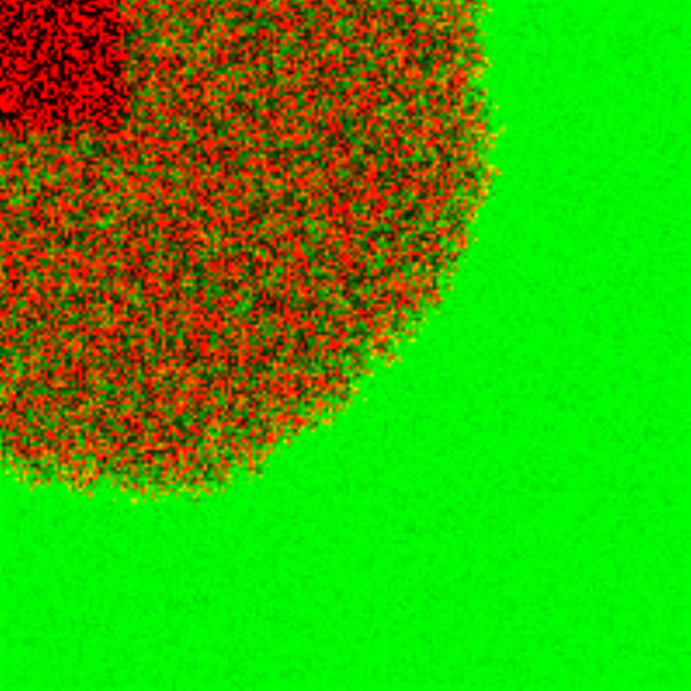}}\hfill
\subfigure[$70$]{\includegraphics[width=0.175\textwidth]{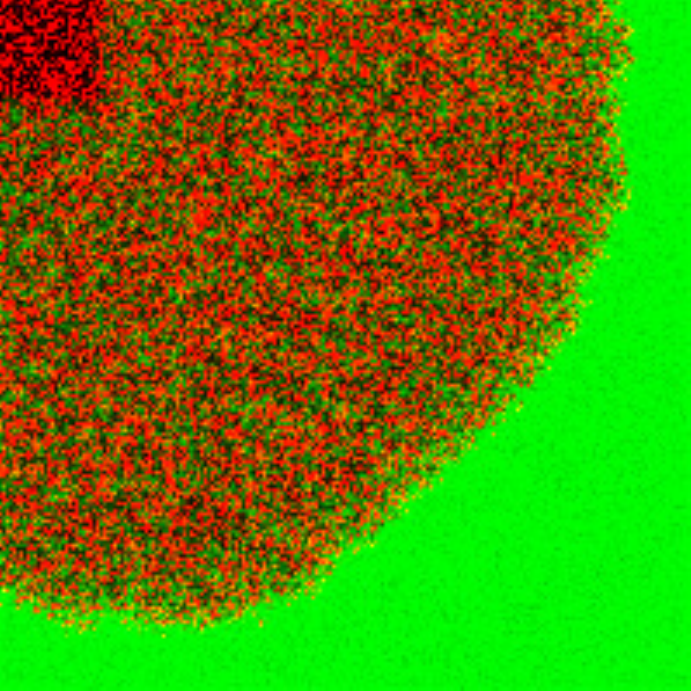}}\hfill
\subfigure[$90$]{\includegraphics[width=0.175\textwidth]{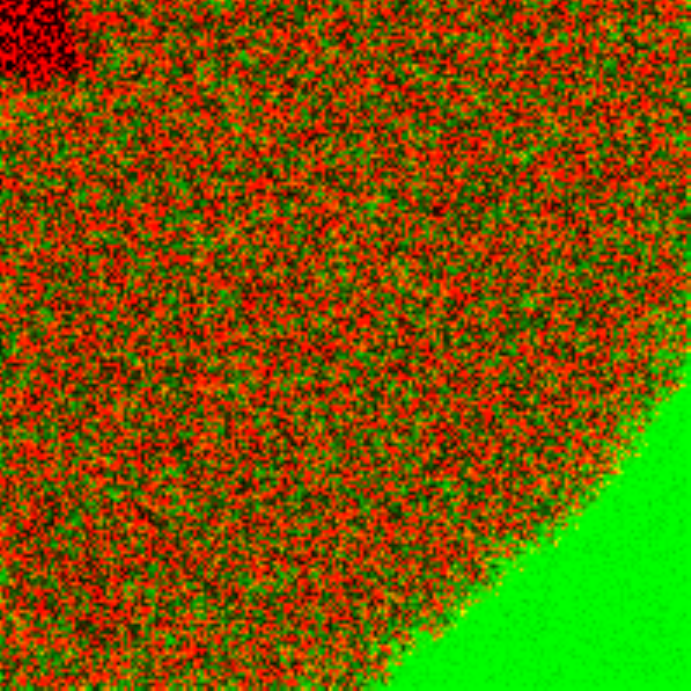}}\hfill
\subfigure[$130$]{\includegraphics[width=0.175\textwidth]{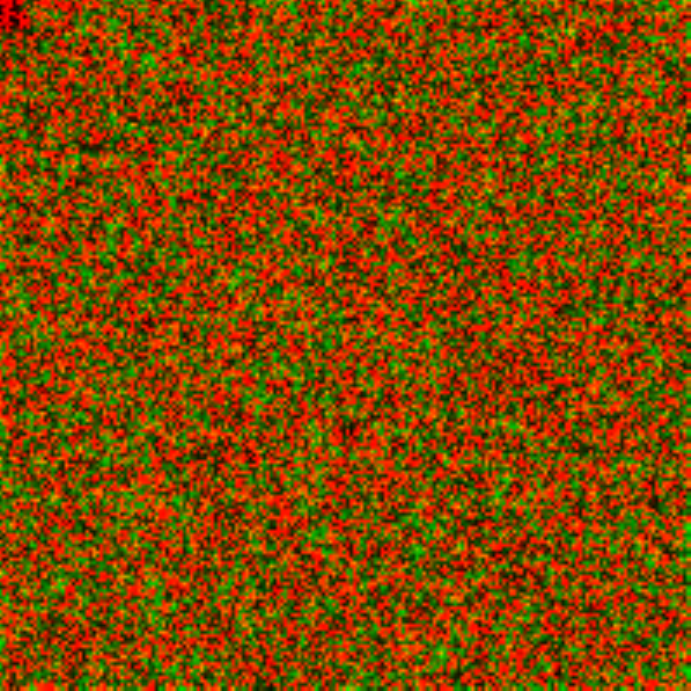}}\hfill
\caption{Rather than the emergence of segregated habitats as in the
  preceding simulations, noise can also mediate coexistence of
  resident and invader across the whole spatial domain. This effect
  does not seem to depend on the model that is chosen for the dispersal
  of populations. Here, we show the pattern for Fickian diffusion}
\label{fig:009}
\end{figure}

The behaviour is similar for Fokker-Planck diffusion but
  here it can be noted that the preference of the resident for certain
  regions of the spatial domain is apparent---within rings less
  favoured by the resident where its diffusivity has maximum values
  the density of the invader is noticeably higher
  (Figure~\ref{fig:010}).  

\begin{figure}[htbp]
\center
\subfigure[$t=10$]{\includegraphics[width=0.175\textwidth]{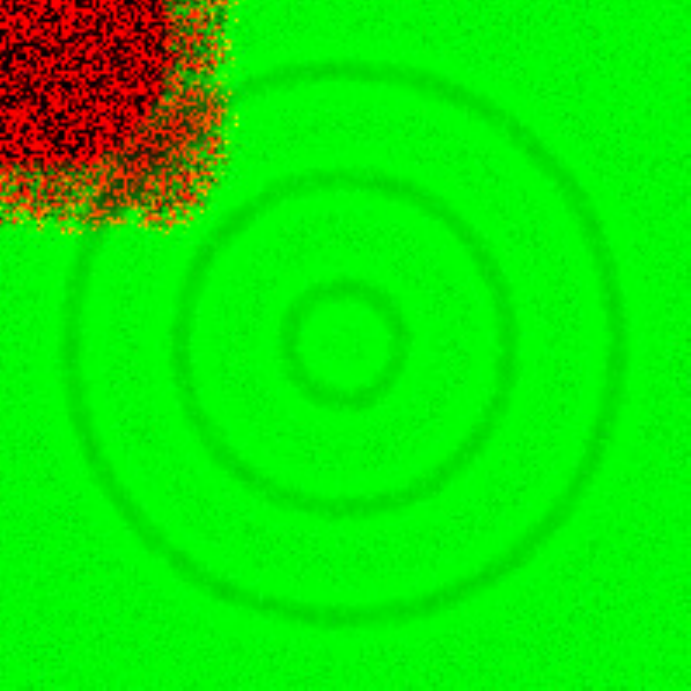}}\hfill
\subfigure[$50$]{\includegraphics[width=0.175\textwidth]{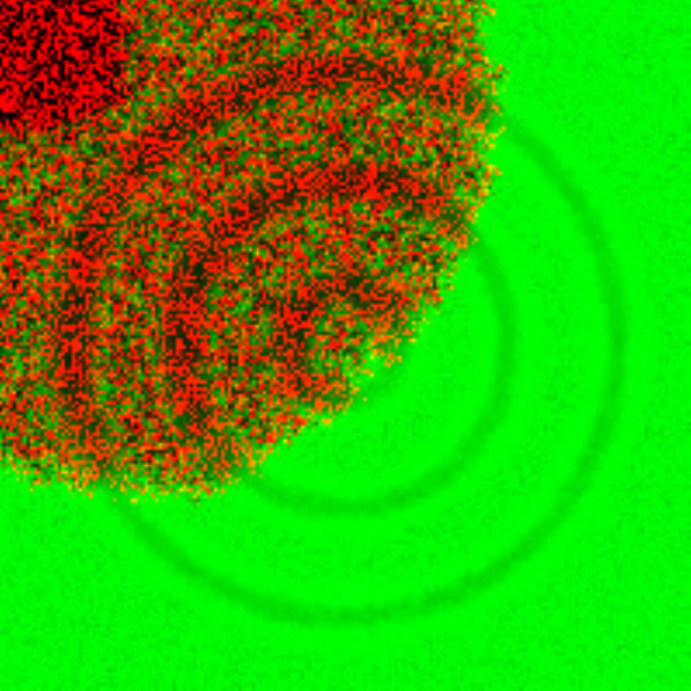}}\hfill
\subfigure[$70$]{\includegraphics[width=0.175\textwidth]{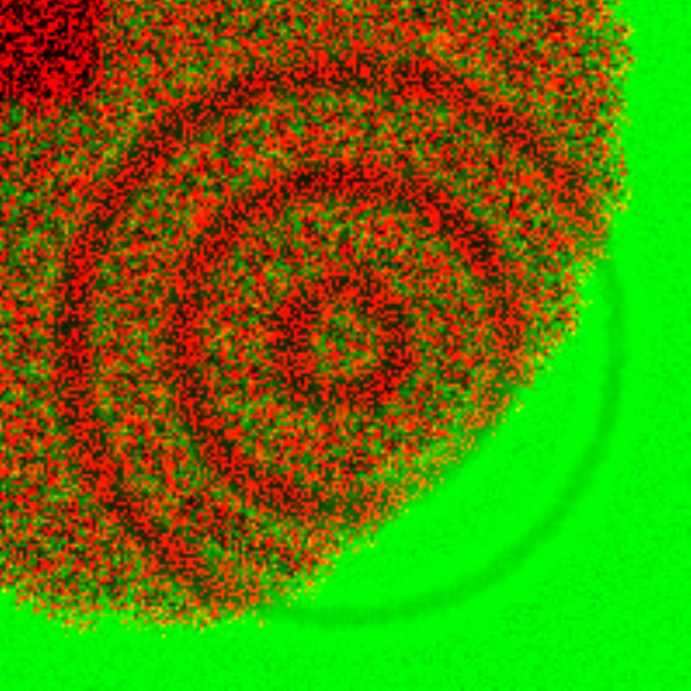}}\hfill
\subfigure[$90$]{\includegraphics[width=0.175\textwidth]{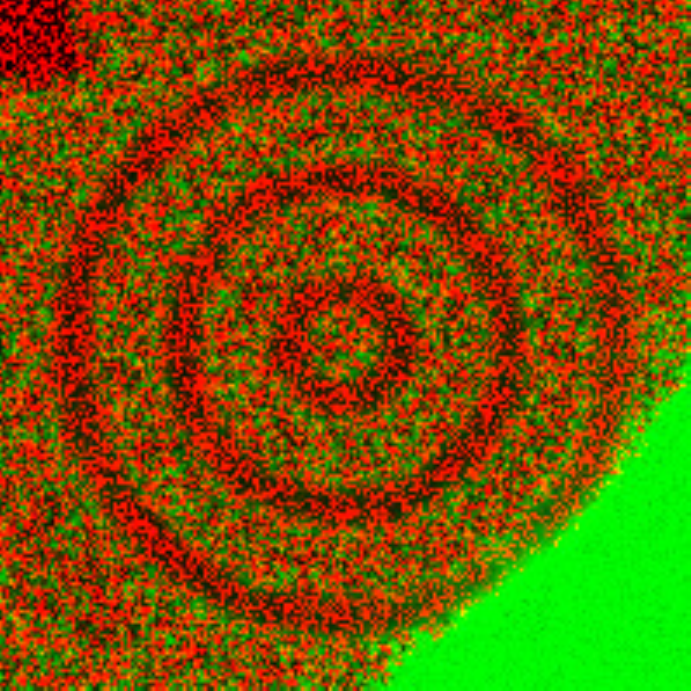}}\hfill
\subfigure[$130$]{\includegraphics[width=0.175\textwidth]{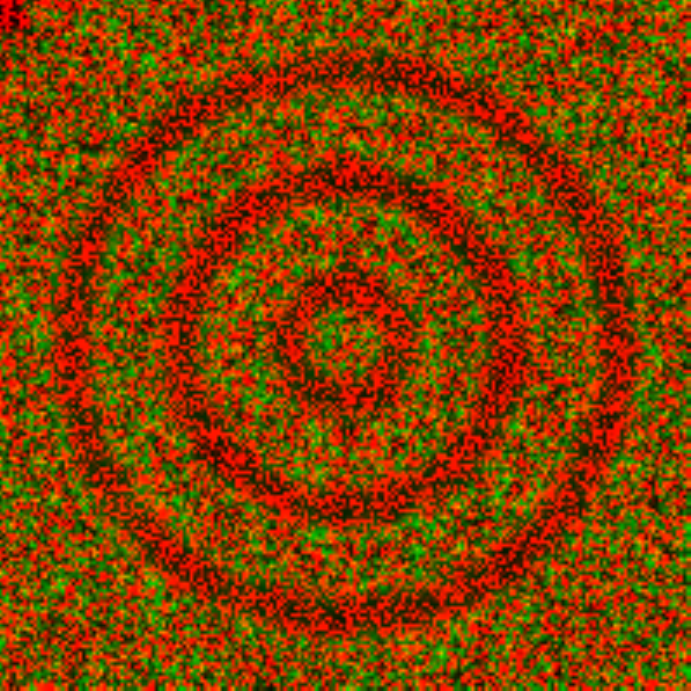}}\hfill
\caption{Coexistence of resident and invader mediated by
  noise. Comparison with Fickian diffusion (Figure~\ref{fig:009}) shows
  that the resident still occupies regions according to its
  preference.}
\label{fig:010}
\end{figure}
}

\section{Conclusions}
\label{sec:conc}

We have studied a biological invasion based on a spatio-temporal
Lotka-Volterra competition model under the influence of stochastic
environmental fluctuations. We found surprisingly rich dynamics after
replacing the two standard models for movement of populations and
environmental variability. Instead of Fickian diffusion, dispersal of
the resident was modelled by the so-called Fokker-Planck law of
diffusion. The most important difference of Fokker-Planck diffusion to
classical Fickian diffusion is that for spatially varying diffusion
coefficients, the stationary distribution is heterogeneous which can
be interpreted as the result of a population's preference for
different spatial regions within their habitat. Moreover, the
Fokker-Planck law of diffusion has a mechanistic basis that is not
less well-suited for modelling movement of biological populations than
Fickian diffusion. Our results show that Fokker-Planck diffusion
facilitates invasion because the invader may manage to establish
itself in spatial areas that are less favoured by the resident.

Similarly, we have considered an alternative for the classical model
for environmental fluctuations based on noise intensities that
increase linearly with the population densities. One interpretation
based on branching processes in a random environment (BPRE) is that
each individual is affected independently by environmental
stochasticity, an assumption that is clearly debatable for large
population densities. Replacing the linearly increasing noise
intensities by a model for which noise intensities saturate or even
decrease with increasing population densities accounts for the fact
that individuals rather than responding independently are similarly
affected by environmental perturbations. The most important
qualitative difference of this model is the fact that the noise
intensity does not exceed a certain upper bound even for large
population densities. Similar to the classical model, increasing the
noise intensity of an invader usually facilitates invasion by
increasing the speed of invasion.

Whereas for many parameter sets the model predicts a segregation of the spatial domain into separate habitats for resident and invader, the new noise model produces an even more interesting effect. Mediated by the noise, resident and competitor are even able to coexist in a mixed habitat that extends across the whole spatial domain. It could be shown that this effect is indeed primarily due to noise because it occurs for both Fokker-Planck as well as Fickian diffusion.

\bibliographystyle{plain}
\bibliography{refer}{}


\end{document}